\documentclass[]{aa}
\usepackage{natbib}
\usepackage{psfig}
\def\nH2{{\rm n}({\rm H}_2)}
\def\NH2{{\rm N}({\rm H}_2)}
\def\pccc{{\rm cm}^{-3}} \def\pcc{{\rm cm}^{-2}}

\def\Tstar#1 {\mbox{${\rm T}_{\rm #1}^*$}}
\def\Tsub#1 {\mbox{$T_{\rm #1}$}}
\def\TK  {\Tsub K }

\def\TD  {\Tsub D }
\def\TL  {\Tsub L }
\def\Tsp {\Tsub sp }

\def\vturb{\mbox{v$_{turb}$}}

\def\p{$^+$}

\def\h13cop{\mbox{{H$^{13}$CO\p}}}

\def\c3h2{\mbox{C$_3$H$_2$}}
 
 \def\R0{R$_0$}

\def\ddeg{{}^\circ\kern-.1em}

\def\kms{\mbox{km\,s$^{-1}$}}
\def\ps{\mbox{s$^{-1}$}}

\def\E#1{\,10^{#1}}
\def\P#1,{$\nH2\TK~=~#1\times~10^4~\pccc$~K}
\def\ec#1,#2,#3,{#1\,(#2)\E{#3}}

\def\H3{\mbox{H$_3$}}
\def\Lya{\mbox{Ly-$\alpha$}}

\def\zcg{\mbox {$\zeta^\gamma_C$}}
\title{The Spin Temperature of Warm Interstellar H I}

\author{H. Liszt\inst{1}}
\institute{National Radio Astronomy Observatory,
           520 Edgemont Road,
           Charlottesville, VA,
           USA 22903-2475}

\begin{document}
\date{received \today}
\offprints{H. S. Liszt}
\mail{hliszt@nrao.edu}

\abstract{
Collisional excitation 
of the $\lambda$21cm hyperfine transition is not strong enough 
to thermalize it in warm neutral (``intercloud'') interstellar gas,
which we show by simultaneously solving the equations of ionization and 
collisional equilibrium under typical conditions.  Coupling of
the $\lambda$21cm excitation temperature and local gas motions
may be established by the \Lya\ radiation field, but only
if strong Galactic \Lya\ radiation permeates the gas in question.
The \Lya\ radiation tends to impart to the gas its own characteristic
temperature, which is determined by the range of gas motions
that occur on the spatial scale of the \Lya\ scattering.
In general, the calculation of H I spin temperatures is a more
difficult and interesting problem than might have been expected,
as is any interpretation of H I spin temperature measurements.
\keywords{ interstellar medium -- H I }
}
\maketitle

\section {Introduction.}

It is an article of faith among H I observers that the spin (excitation)
temperature \Tsp\ of the ground-state $\lambda$21cm hyperfine transition
is equal to the kinetic temperature of the ambient gas.  
This belief of course arises from the weakness of the transition,
the consequent long lifetime against spontaneous emission 
(A$_{21} = 2.85\times10^{-15}$ \ps\ for H I), and the
easily-demonstrated dominance of particle collisions with other H-atoms
in cool gas. Much of our basic 
understanding of the phases of the ISM derives directly from measurements 
of the H I spin temperature, for instance by comparing nearby or
overlapping absorption and emission profiles \citep{DicTer+78,PayTer+82}.

But spin temperature measurements often seem to imply unphysical kinetic 
temperatures: not wildly so, say negative or infinite, but definitely 
in a range -- 1000 to 5000 K -- where the interstellar gas is unstable in 
multi-phase models \citep{WolHol+95,McKOst77} and therefore 
should be so short-lived as to be unobservable.  One might be tempted to 
disparage either theory or the data but \cite{DavCum75} took a somewhat 
less doctrinaire view and showed that, for some assumed combinations of 
conditions chosen to be representative of gas which might produce pulsar 
dispersion measures, the density of collision partners in warm or intercloud 
gas is simply too small to thermalize the line.  Although little attention
seems to have been paid to this warning, it suggests that we remain open 
to the possibility that blind faith in the equality of the spin 
and kinetic temperatures might be misplaced.

Calculation of the H I spin temperature turns out to be a remarkably
complex problem, involving intimate knowledge of the ionization and
phase structure of the ISM, as well as its topology.  Here we update and 
expand upon the discussion of \cite{DavCum75} by actually calculating
the ionization and collisional excitation in interstellar H I regions,
in the process demonstrating the inability of particle collisions to thermalize
the $\lambda$21 cm transition.  But we also consider an important
mechanism which \cite{DavCum75} ignored, whereby the \Lya\ radiation field
threading and partly produced by the gas tends to impart to the $\lambda$21 
cm transition its own effective temperature.  \Lya\ photons acquire this 
temperature while undergoing large numbers (10$^7$ or more) of repeated 
scatterings on many spatial scales ranging upward from the line-center
mean free path, $0.11$ (\TD/100K)$^{0.5}/$n$_{\rm H}$ AU (\TD\ is the
Doppler temperature), and it is therefore representative of the motions 
(thermal, turbulent, $etc.$) which are established in the gas on those 
scales.  

That excitation by light dominates hyperfine excitation locally around 
individual stars is well known in the context of scattering of Solar 
\Lya\ radiation: \cite{BraKyr98}, for instance, point out that Solar 
\Lya\ radiation dominates in local intercloud H I within 1000 AU of 
the Sun, and that the sphere of influence of an O-star would necessarily 
be much larger.  In the interstellar context, we will
refer to excitation of the hyperfine line by scattered \Lya\ radiation as the 
Wouthuysen-Field mechanism or WF effect following the discussions in 
\cite{Wou52} and \cite{Fie58,Fie59c} and we find that it 
opens up interesting possibilities for the spin temperature.  At low density 
in the intercloud medium, warm neutral H I may (or may not) be Dopplerized by 
Galactic \Lya\ photons but is certainly {\it not} thermalized by local 
particle collisions.  In general, the importance which is assigned to \Lya\ 
excitation in intercloud gas depends directly on the topology and large-scale 
structure which is assumed to apply in the interstellar medium.

The plan of the current discussion is as follows. In Section 2 we show
what is needed to calculate the excitation of the hyperfine transition.   
In Section 3 we lay out the basics of the ionization equilibrium calculations 
which are needed to determine the densities of the main collision partners 
(electrons and neutral H-atoms) under given conditions of local gas density 
and gas thermal pressure.  These calculations also serve 
to determine the locally-produced field of \Lya\ photons.  In Section 4 
we discuss results of the combined ionization and particle excitation 
calculations. In Section 5 we introduce the behaviour of the internally-
generated \Lya\ radiation field, and conclude that it also is not strong
enough to influence the hyperfine excitation in neutral gas.  In Section
6 we discuss the effect of the much stronger galactic \Lya\ photon
field on the H I in various multiphase models of the ISM.

\section{Hyperfine excitation}

Discussions of the excitation of the $\lambda$21cm transition generally
follow that of \cite{Fie58} who distinguished among three types of
excitation; radiative, by a radiation field around the rest wavelength of the 
hyperfine transition having a characteristic brightness temperature 
\Tsub R \ as seen in the gas (\Tsub R \ = 2.73K here);
collisional with a total downward rate R$^c_{21}$ (\ps) at kinetic
temperature \Tsub K \ ; and radiative in the \Lya\ photon field, with a 
net downward rate R$^\alpha_{21}$, at an effective temperature \Tsub L \ .  
Field derived an expression for the \Lya\ excitation which was used in an
equivalent form by \cite{BahEke69} and which will be discussed below.

In these terms the spin ({\it i.e.} excitation) temperature of the 
$\lambda$21cm line can be expressed as

$$ \Tsp = (\Tsub R + y_c + y_\alpha)/(1+y_c/\TK + y_\alpha/\TL) \eqno(1) $$

with

$$ y_c = T_{21} R^c_{21}/A_{21}, $$
$$ y_\alpha = T_{21} R^\alpha_{21}/A_{21} $$

and where T$_{21} \equiv h\nu_{21}/k = 0.068$ K, 
A$_{21} = 2.85 \times 10^{-15}\ps$.

Equation 1 shows that R$_{21} >> 15 \TK A_{21}$ is required in order to 
thermalize the transition.

\subsection{Collisional excitation by electrons, protons, and neutral H-atoms}

The required rate R$^c_{21}$ is the sum of the
downward collision rates over all collision partners (H-atoms, protons,
and electrons).  The downward rate constant for interactions with electrons 
is taken from \cite{Smi66}:  we note that is well-represented by the 
functional form 

$$\log\gamma^e_{21}(\TK)=-9.607+\log(\sqrt{\TK})\exp(-(\log \TK)^{4.5}/1800) $$

for \TK $\leq$ 10$^4$ K and

$$ \gamma^e_{21}(\TK \geq 10^4~K) = \gamma^e_{21}(10^4~K) $$

over the range considered here (the logs are base 10 and the units of
$\gamma^e_{21}$ are cm$^3$\ps)

\cite{Smi66} also gives the rate coefficient for proton de-excitation, which
is just 3.2 times larger than that for neutral atoms at \TK $>$ 30 K.  We used
$\gamma^p_{21} = 3.2 \gamma^H_{21}$ but employed the more recent neutral atom 
de-excitation rates of \cite{AllDal69} which we fit in piece-wise 
continuous fashion, following their tabular presentation.  
Excitation by protons is unimportant because it is so much weaker than 
that by electrons at the same temperature and because n$_e > {\rm n}_p$ under 
all conditions.  

The results of \cite{AllDal69} are given up to 1000 K at which point 
they are increasing as \TK$^{0.33}$, but with a negative second derivative. 
The neutral-atom rate coefficients of \cite{Smi66} increase somewhat 
faster for \TK\ $>$ 1000 K with a slightly positive second derivative,
but are not substantially different in the mean.  We 
extrapolated the rate coefficients of \cite{AllDal69} to higher 
temperature as \TK$^{0.33}$. In this regime the excitation 
is increasingly dominated by electrons and the results reported here
are not noticeably influenced by the manner of the extrapolation.

\subsection{Excitation by the \Lya\ photon field}

Recoil -- momentum conservation -- at each scattering causes the field of 
\Lya\ photons to have a slope (color temperature) at the line center which 
corresponds to the Doppler temperature of the ambient medium
\citep{Fie59c,Ada71a}; \Lya\ photons sample the gas kinematics and then 
tend to impart to the hyperfine transition a corresponding excitation 
temperature.  \cite{BahEke69} and \cite{UrbWol81} employed this mechanism 
to show how the H I spin temperature could remain low, rendering 
$\lambda$21cm absorption detectable, even in neutral gas near quasars.
In cases where the gas kinematics are affected by non-thermal motion, 
this will be reflected in the spin temperature, depending on the relative 
importance of collisional and \Lya\ excitation. 

\cite{Fie58} considered detailed balance in the various transitions
between sublevels which constitute the \Lya\ line and arrived at the
following expression for y$_\alpha$ 

$$ y_\alpha = (c T_{21}/36\pi) (m_{\rm H}/k\TD)^{0.5} 
\lambda^{3}_\alpha (A_\alpha/A_{21}) n_\alpha \eqno(2) $$

where $A_\alpha$ is the spontaneous emission rate 
($4.7\times10^8$ \ps\ corresponds to the mean \Lya\ absorption f-value 
f=0.4167), n$_\alpha$ is the number density of \Lya\ photons at the 
line center and \TD, the Doppler temperature, represents the
width of the Gaussian core of the line.  \TD\ is defined in the
usual way \citep{LeuLis76} by the expression

$$2k\TD/m_H = 2k\TK/m_H + \vturb^2 \eqno(3) $$

leading to \TD\ = \TK+121 K (\vturb/1 \kms)$^2$.  This expression
accords with the definition used by \cite{PayTer+82} \TD\ = 21.86K W$^2$ 
where W is the FWHM (\kms) and differs from the earlier usage of 
\cite{DicTer+78}, \TD\ = 121K W$^2$ taking W as the HWHM.  The Doppler 
temperatures shown by \cite{DicTer+78} are actually too large by a factor 
2*ln(2) $\approx$ 1.4.

\begin{figure}
\psfig{figure=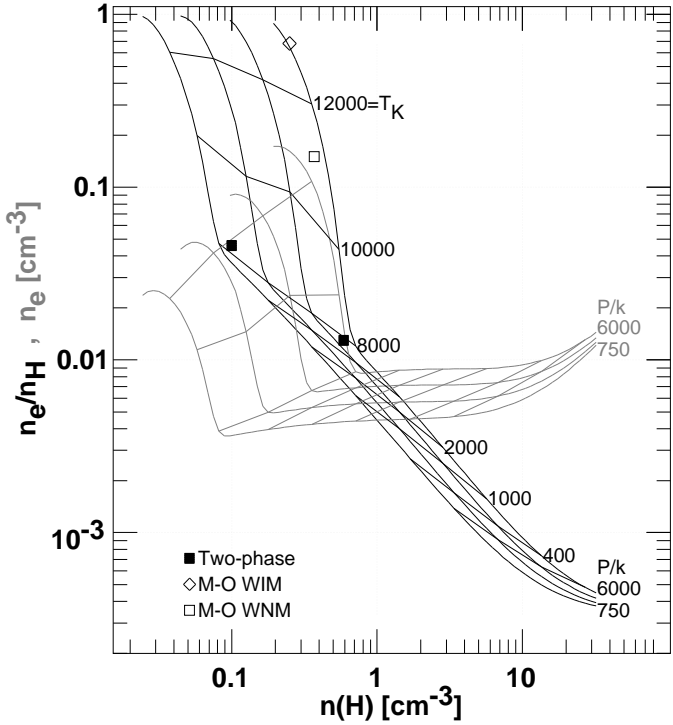,height=9.5cm}
\caption[]{Ionization equilibrium calculations in atomic gas.
The full curves are plots of the ionization fraction 
n$_e$/n$_{\rm H}$ where n$_{\rm H}$ is the number density of H-nuclei in all 
forms.  Results are shown for four values of the total gas pressure 
$P/k = 750, 1500, 3000, 6000~\pccc$ K and the kinetic
temperature is traced across each set of curves.  The shaded curves are
plots of the electron {\it density}. 
Open symbols represent the ionization fraction in warm neutral and warm
ionized gas in the three-phase model of \cite{McKOst77}.  The filled
rectangles show the ionization fraction of warm neutral gas in the two-phase
calculations of \cite{WolHol+95} for their standard case Nw $= 10^{19}~\pcc$.}
\end{figure}

Evaluating the constants in equation (2) we find

$$ y_\alpha = 5.90\times10^{11} {\rm n}_\alpha \TD^{-0.5}. $$

\section{Ionization equilibrium in H I regions}

\subsection{Calculations}

To obtain number densities of the various collision partners which 
determine the equilibrium of the hyperfine transition it is necessary 
to calculate the ionization equilibrium of a dilute gas.  We duplicated 
the relevant portions of the work of \cite{WolHol+95}
which is a very comprehensive study of two-phase equilibrium
in the interstellar medium and must be regarded as the standard
reference on this subject.  That is, we consider a gas consisting 
of hydrogen, helium and carbon, ionized by soft x-rays and cosmic
rays (and their secondary electrons) which recombines in both the gas 
phase and on small grains (following the formalism of \cite{DraSut87}).   
This calculation is fully described in the original reference and
a detailed discussion will not be repeated here. We 
note that it uses Solar metallicity, no appreciable gas-phase 
depletion of carbon ([C]/[H]$=3.3\times10^{-4}$), and a primary cosmic 
ray ionization rate $\zeta^{cr}_H = 1.7\times10^{-17}$ \ps.
The ionization fractions are overestimates by factors of 2-3 in 
higher-density gas if gas-phase depletion is important.

\cite{WolHol+95} parametrized their model in terms of a column density of 
hydrogen nuclei Nw which attenuates part of the soft x-ray flux; their 
standard model takes a value Nw = $10^{19}~\pcc$ for this quantity,
with factor of ten variations in either sense as subsidiary examples.
To model the x-ray ionization rate, we calculated the total attenuation
at each energy over a gas column Nw (due to all the important elements,
not just those whose ionization is calculated)
using the absorption cross-sections of \cite{BalMcC+92}
and modulated the free-space x-ray spectrum accordingly. The product
of this attenuated and modulated x-ray spectrum and the absorption
cross section is then integrated over energy to give the primary
ionization rate for hydrogen and  helium.  We also calculated the 
mean photoelectron energy for use in determining the secondary ionization; 
this is typically 40-50 eV as for cosmic-ray events.

\subsection{Recombination coefficients}

\begin{table}
\caption[]{Recombination rate coefficients}
{
\begin{tabular}{lcc}
\hline
Species& a (cm$^3$ \ps) & b \\

\hline
H$_{n>1}$   &$3.50\times10^{-12}$ & 0.75 \\
H$_{^2S}$   &$1.54\times10^{-13}$ & 0.54$^a$ \\ 
He          &$2.36\times10^{-12}$ & 0.64 \\
C           &$4.67\times10^{-12}$ & 0.60 \\
\hline
\end{tabular}}
\\
$^a$ Rate into the $^2S$ level only, see Section 3.2 \\
\end{table}

We took recombination coefficients for various species from the 1999 
update of the UMIST reaction database.  They are expressed in the form 
$\alpha_Y = a_Y\times(\TK/300)^{-b_Y}$ where the coefficients are given in 
Table 1.  Hydrogen of course is a special case whereby the needed 
recombination coefficient is that into the n=2 state and higher, 
since recombinations into n=1 produce ionizing photons which are 
immediately reabsorbed in the surrounding gas.  A special case is 
considered below whereby we require the recombination rate for 
production of \Lya\ radiation.  In this case, recombinations into the
metastable 2S state must be ignored, and the effective recombination 
rate is diminished by an amount which is given in the second entry
of the Table, taken from \cite{Mar88}.

\subsection{Photoionization of carbon}

For this we take from the UMIST database the free space value
\zcg(0) = $3.0\times10^{-10}$ \ps.  This can then be traced into 
a gas column as \zcg(A$_V$)= \zcg(0)exp($-2.42A_V$) following 
\cite{BlaDal77} but we will consider only the case $A_V = 0$.  The
major contribution to the ionization fraction for the warm gas considered
here is generally from H and He, not from carbon.

\begin{figure}
\psfig{figure=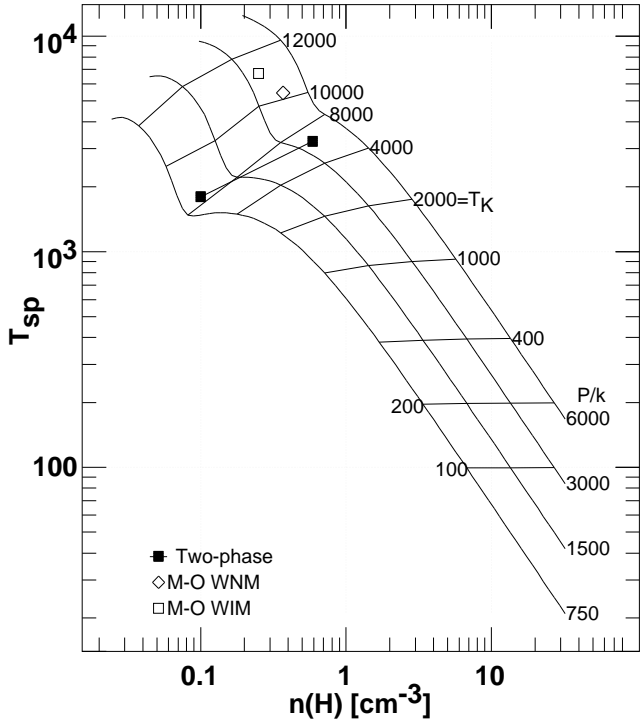,height=9.5cm}
\caption[]{The spin temperature of the H I $\lambda$21cm transition
resulting from the same conditions whose ionization equilibrium is 
shown in Figure 1.  Symbols represent results for warm gas in
the two and three phase calculations of \cite{WolHol+95} and \cite{McKOst77}, 
see Figure 1.}
\end{figure}

\subsection{Collisional ionization of hydrogen by electrons}

We used the electron-impact collisional ionization rates of 
\cite{SchWal91}.  The reader is cautioned to ignore an unfortunate 
10 order of magnitude error in labelling of the vertical scale of their 
Figure 1, lest this crucial process be ignored.

\section{Ionization equilibrium and hyperfine excitation
 in the interstellar hydrogen}

\subsection{The ionization fraction as a function of gas pressure and 
density}

In Figure 1 we show the results of our ionization calculations for four values 
of the gas thermal pressure, using the standard value Nw $= 10^{19}~\pcc$. 
The ionization fraction is shown as solid lines and the electron density 
is shown shaded.  The kinetic temperature is traced across the set of four 
curves of varying pressure.  Clearly, the ionization fraction in neutral gas 
is controlled by the total density, while the electron density -- which 
varies remarkably little with $n_{\rm H}$ until the gas ionizes fully -- is 
influenced somewhat by the {\it pressure}.  The soft x-ray ionization rate 
scales the overall behaviour, albeit somewhat weakly (not shown).

The filled dark squares in Figure 1 represent the limiting cases at high and 
low pressure for the model of two-phase equilibrium from \cite{WolHol+95}.  
Table 2 gives the range of pressure and kinetic temperature over which two
phases can coexist for various values of Nw and it shows that two-phase 
equilibrium is possible over a wider range in \TK \ than is sometimes 
considered when interpreting spin temperature measurements.  Comparison of 
the Table
entries for the standard case Nw $= 1\times10^{19}~\pcc$ with the locations
of the filled dark symbols in the Figure shows that our results mimic those of 
\cite{WolHol+95}, as intended.  Both sets of calculations produce the same 
ionization at any given density and temperature.

In Figure 1, two unfilled symbols represent the results of \cite{McKOst77} 
for the warm neutral and warm ionized gas (both at 8000K).  The pressure 
in the McKee-Ostriker model is P/k = 3700 K $\pccc$ but the ionization
fraction in three-phase warm neutral and ionized gas is considerably
higher than that in our calculations. \cite{McKOst77} explain 
in the caption to their Figure 1 that the warm neutral medium can only be 
produced in appreciable quantity at the assumed pressure by assuming a higher 
than average value of the soft x-ray flux.  In the context of our
modelling, the soft x-ray flux would have to be increased by a factor
of about 50 in order to give the same total ionization which occurs in 
warm, neutral gas in the McKee-Ostriker model.  It is unfortunate that
the basic parameters of the three-phase model have not been revised
in so long.

\begin{table}
\caption[]{Two-phase spin and kinetic temperatures}
{
\begin{tabular}{lccc}
\hline
Model & P/k (cm$^3$ K) & \TK (K) & \Tsp (K)\\
\hline
Nw=$10^{18}~\pcc$ & 1600 - 9990 & 9200 - 5600 & 3525 - 4620\\
Nw=$10^{19}~\pcc$ & 990 - 3600 & 8700 - 5500 &1800 - 3260 \\
Nw=$10^{20}~\pcc$  & 610 - 1500 & 8200 - 4900 & 1035 - 2020 \\
Three-Phase & 3700 & 8000 & 5400 \\
\hline
\end{tabular}}
\end{table}

\subsection{Evaluation of \Tsp}

Figure 2 shows that for kinetic temperatures above 2000 K, and especially
at 8000 K, the particle excitation rate is too small to thermalize the 
$\lambda$ 21cm line in any model of the ISM, including the McKee-Ostriker
three-phase picture.  The extreme conditions assumed for the three-
phase model result in \Tsp $\approx 5400$ K in warm neutral gas with
\TK = 8000 K.  In the two-phase model, the spin temperature of warm gas 
would be smaller, 1800 - 3200 K for the standard conditions where
\TK = 8700 - 5500 K (the models at higher \TK\ have lower pressure
and lower \Tsp). Table 2 shows the range of spin temperature which 
results from various conditions of two- and three-phase equilibrium.   
Figure 3 displays the y-factor for collisional excitation corresponding 
to the conditions shown in Figures 1 and 2.  In the warm gas, 
$y_c/\TK \approx 1$, while $y_c/\TK >> 1$ is the relevant condition for 
thermalization.

Physically, the subthermal excitation comes about because the condition 
for thermalization becomes ever more stringent at higher kinetic 
temperature, while the neutral particle excitation rate coefficient only 
increases as \TK$^{0.3}$ \citep{AllDal69}.  The neutral particle 
density declines as 1/\TK\ at constant pressure until the gas begins 
to ionize, at which point new particles enter the gas and the decline 
is even more rapid.  Once the gas ionizes even slightly (a few percent), 
excitation by electrons begins to dominate and 
the remaining neutral atoms will be more strongly excited.  But the 
transition from neutral to ionized gas (which increasingly precludes 
observation of the $\lambda$21cm line) occurs at too low a density to 
preserve the thermalization of the line even when a very large soft x-ray 
flux is asserted, as in the model of \cite{McKOst77} 
\footnote{Perhaps this model might also be referred to as the Lake Woebegon
solution, since it seems to assume that the soft xray field is everywhere 
brighter than average.}.

We conclude that in the context of these calculations, one simply cannot 
expect to observe spin temperatures which are directly indicative of the 
8000 K kinetic temperature typically assumed (by observers) for warm gas.  
Instead, seemingly unphysical, intermediate values will appear, even
when the contribution of warm gas can be isolated. This
situation comes about mostly because the excitation is sub-thermal, but
also because multi-phase equilbrium is not entirely oblivious of local
conditions; the kinetic temperature of warm neutral gas 
can vary between 5000 and 9000 K in two-phase models depending on the 
pressure, elemental abundances, {\it etc.}  Moreover,
\cite{WolHol+95} note that the time to reach thermal equilibrium
in warm gas is too long to ensure that a steady-state actually
occurs; rather, two-phase equilibrium is an ideal, toward which the 
ISM may tend.
If seemingly forbidden spin temperatures 1000-5000 K are actually measured 
in the $\lambda$21cm line, it is unfair and improper to impugn 
the multi-phase models on this account, because they predict them.

Finally, we note that the need to consider a wide range of spin temperatures
in warm gas makes it harder to discern \Tsp\ in cool gas when both are 
seen blended along the line of sight.  Of course, lines of sight in which
the contribution of cool gas is {\it not} strongly blended are almost
unknown.  The presence of lower \Tsp\ in the intercloud medium would
result in lower \Tsp\ in line blends, perhaps producing another range
of seemingly unphysical \Tsp\ between $~\approx 1000$ K and the range 
expected for cool gas alone, 40-200 K \citep{WolHol+95}.

\section{Excitation by the \Lya\ photon field}

In the previous section we showed that the $\lambda21$cm hyperfine
line will not be thermalized by particle collisions in warm weakly ionized gas.
But there is another means of excitation available, the Wouthuysen-Field
effect, which we now discuss.

\subsection{\Lya\ scattering}

Our basic understanding of the behaviour of \Lya\ photons in very optically 
thick media dates from the discussion of \cite{Fie59c} on relaxation in
Doppler broadening and the numerical work of \cite{Ada72} on scattering 
including strong damping.  The latter showed that 
the mean number of scatterings $\langle N \rangle$ of a photon before
escape from a lossless medium never exhibits the behaviour expected of 
a pure one-dimensional random walk.  Taking $\tau_0$ as the line-center 
optical depth at the midplane of a semi-infinite slab viewed along the 
outward normal, Adams found instead that, at sufficiently high
optical depth (see \cite{HumKun80}), $\langle N \rangle \approx \tau_0$.  
This seems to have been unexpected, even though identical behaviour 
is discussed by \cite{Fie59c}, citing \cite{Zan49}, following essentially
the same line of reasoning.

\begin{figure}
\psfig{figure=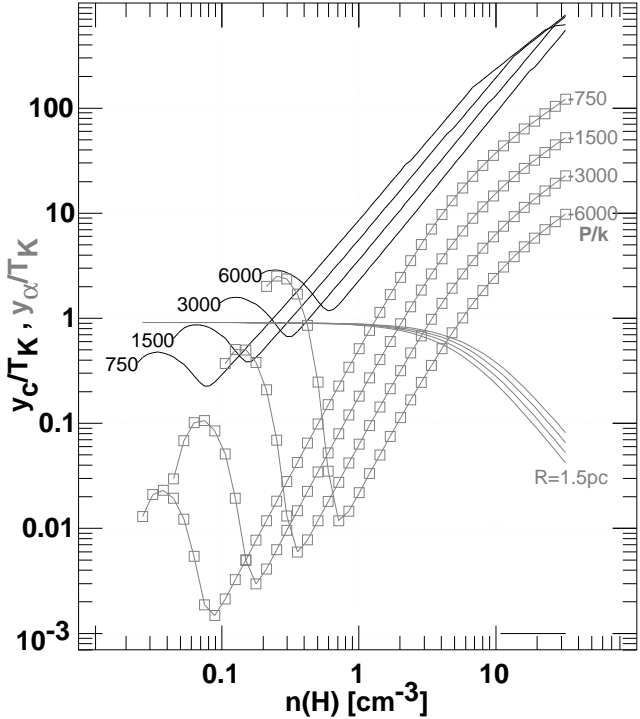,height=9.5cm}
\caption[]{The y-parameter for collisional (solid curves) and 
\Lya\ excitation (shaded and chained curves) for a medium of assumed
size 2R = 3 pc.  The same 4 thermal pressures are employed here as in 
Figures 1 and 2.   The shaded curves marked by ``R=1.5pc''
represent the correction factors, accounting for radiative transfer in
lossy media, which have been applied to the \Lya\ excitation following
the discussion in Section 5.2; see also the following Figure.
The horizontal line at lower right shows where 
N$_{\rm H}$ = 2n$_{\rm H}$R $= 1-3\times10^{20}~\pcc$.}
\end{figure}

\cite{Ada72} explained this result heuristically by noting another aspect 
of the simulations, namely that photons escape the medium only in the far 
line wings of the Voigt profile $\phi(x)$, where x is a normalized 
frequency displacement from line center just below.  They escape at 
$|x_\ast| \equiv |(\nu_\ast-\nu_0)|/\Delta\nu_D \approx\ (a\tau_0)^{1/3}$ 
where $\Delta\nu_D = (\nu_0/c)\sqrt{2ln(2)k\TD/m_H}$ is the Doppler HWHM and 
a is the damping constant measured in units of $\Delta\nu_D$.  
Although most scattering occurs nearer the Doppler core, photons move 
appreciably in space only when they scatter at frequencies where the medium 
is less opaque.  The photons escape at a frequency where the optical
depth is still well above unity; 
$|x_\ast|/|x_{\tau=1}| = 0.285(\TD/10^4 K)^{1/6}(N_H/10^{20}~\pcc)^{-1/6}$
where $\pm x_{\tau=1}$ are the frequency shifts at which the optical
depth drops to unity for a pure damping profile.

Given that the rms Dopper shift in any scattering is $|\Delta\nu_D|$
(the mean shift is 0 in the core and $-1/|x|$ in the far wings),
Adams reasoned that, occasionally, a photon would experience
one series of scatterings into the far wings carrying it at once 
$|x_\ast|$ frequency units from the line center and  entirely out of
the scattering medium.
If a photon must travel a distance $\tau_0$ to 
escape, $x_\ast$ should satisfy the condition $x_\ast/\phi(x_\ast)=\tau_0$
since the scattering length is proportional to $1/\phi(x)$.
This recovers the behaviour seen in the simulations since 
$\phi(x) \propto a/x^2$ in the damping wings.

Shortly after this work, the transfer equation was effectively
solved in closed form by \cite{Har73} in the case of very large
optical depths $(a\tau_0)^{1/3} > 1$.  The behaviour seen
in the simulations of \cite{Ada72} was recovered analytically in 
the change of variables needed to effect the solution: to
something which varies as x$^3$ in the line wings.  The numerical
work was extended by \cite{Ada75} and \cite{HumKun80}, and
the analytic work by \cite{Neu90,Neu91}.  There, it was
shown that the mean distance travelled by a \Lya\ photon is actually 
$(a\tau_0)^{1/3}\tau_0$.  Because $a = 0.0425/\TD^{0.5}$ is 
typically a small number, the cube root factor is of order 10-100.  
This description is correct to factors of order unity, as discussed
by \cite{HumKun80}.

For reference, we note that the line center optical depth
$\tau_0 = 5.90\times10^{-12} N_{\rm H}/\TD^{0.5}$ is proportional to 
A$_\alpha/\Delta\nu_D$, which is the damping constant a except for a 
numerical factor.  Thus $a\tau_0 = 1.39\times10^{-10}a^2 N_{\rm H}
=  2.51 \times10^{-13}N_{\rm H}/\TD$.  If $\TD \approx \TK$, the 
latter ratio of column density to temperature is familiar from the 
expression for the integrated optical depth of the hyperfine line.  
Thus we come full circle and find (but do not intend to claim a causal 
connection!) that the crucial parameter for \Lya\ excitation of the 
$\lambda$21cm line is proportional to the $\lambda$21cm optical depth.  
\footnote{The discussion of the previous two paragraphs neglects 
internal absorption, which is discussed in detail in Section 5.3}

\subsection{Internal sources of \Lya\ photons}

We included three locally generated sources of \Lya\ photons in H I
gas:  electron impact excitation of the $^2$P level using the rate constant
calculated by \cite{SchWal91}; electron-proton recombinations into levels 
n $\geq 2$ (except $^2$S) for which the relevant coefficients are summarized 
in Table 1 and for which our rates are very slight upper limits; 
and excitation by the 
same cosmic- and x-ray fluxes which  ionize the gas.  For the latter, 
we follow \cite{DraSal78} and assume that 1.5 \Lya\ photons are generated 
per primary ionization event.  

In general, electron-impact excitation is unimportant except above 
10$^4$ K and cosmic-ray and x-ray generated photons dominate in the 
denser gas.

\begin{figure}
\psfig{figure=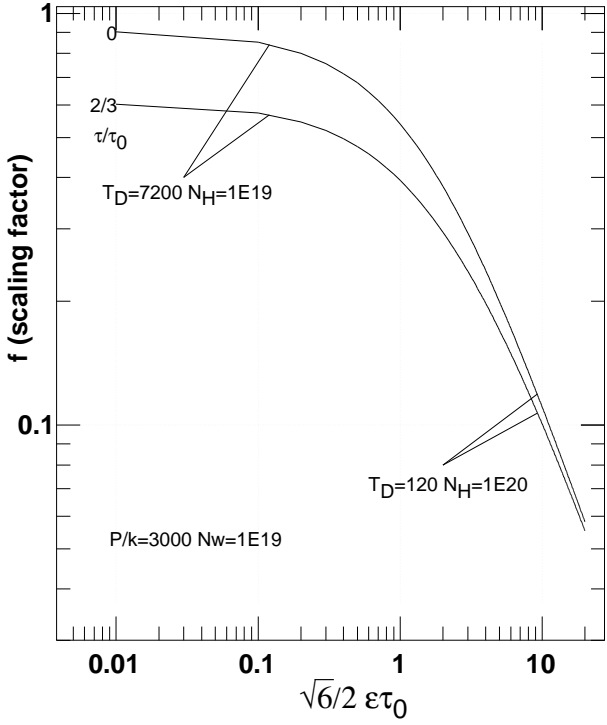,height=9.5cm}
\caption[]{A scaling factor for the mean intensity inside a lossy slab 
harboring a uniform source of \Lya\ radiation, following equations 27 and
41 of \cite{Har73}; see Section 5.3 here.  Values are shown at the 
midplane ($\tau/\tau_0 = 0$) 
and two-thirds of the way from the midplane to the slab surface.  Relevant
values for warm and cool neutral gas are indicated.  These calculations
are the origin of the correction factor displayed in the prior Figure.}
\end{figure}

\begin{figure}
\psfig{figure=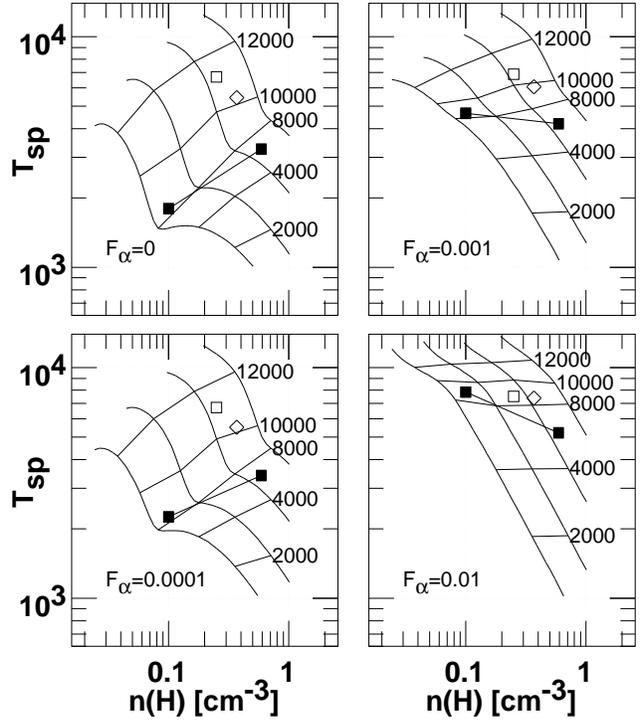,height=9.5cm}
\caption[]{As in Figure 2 when a specified fraction F$_{\alpha}$ of 
the galactic flux from early-type stars threads the gas and \TL = \TK .  
The panel with F$_{\alpha} = 0$ 
repeats behaviour already shown in Figure 2.}
\end{figure}

\subsection{The locally-generated \Lya\ photon field in a dusty H I cloud}

The volume rate at which \Lya\ photons are created, $\dot n_{\alpha}$,
is straightforward to specify. But for the excitation rate we require 
in Field's formulation the volume density of photons 
n$_\alpha = \dot n_{\alpha} \times t$,
where t is a characteristic retention time (the expression could
just as well have been cast in terms of the local mean intensity).
In fact, following \cite{Fie58}, the relevant time is 
R/c where 2R is the characteristic size of the medium.  Although this 
seems naive, it is supported by the analytic solutions of 
\cite{Har73} which we now discuss:  the actual distance 
travelled by a photon, $(a\tau_0)^{1/3}\tau_0$,  is 
associated with a broadening of the profile by an amount 
$(a\tau_0)^{1/3}$, leaving the line-center \Lya\ intensity unchanged.

\cite{Har73} presents an analytic solution for the case of a
uniform slab with a uniform internal source of \Lya\ radiation
$\dot n_{\alpha}$ such that 1 photon is created per unit area per
unit time looking down through a slab of half-thickness R.  To 
first approximation, the mean intensity at the slab midplane, assuming 
a lossless medium, is  
$J(\tau=0,x) = (\sqrt6/2\pi^3){\rm exp}(-1.34|x|^3/(a\tau_0))$. 
To convert this normalized solution to ``real'' units, the density of
photons at the line center is just 
$4\pi/c \times 2R \times J(0,0) = (4\sqrt6/\pi^2) \times (R/c) = 0.993 R/c
\equiv f \times R/c$.

A slightly more accurate expression for the photon field is given by
\cite{Har73} as a series which includes the possibility of accounting for 
losses by continuous absorption in the medium. To calculate the loss it is 
necessary to evaluate the term $(\sqrt{6}/2)\epsilon\tau_0$ where $\epsilon$ 
is the (assumed small) probability of absorption per scattering (recall that
there are on average $\tau_0$ scatterings).  The mean distance travelled per 
scattering, measured in units of the distance necessary to traverse one 
optical depth at the line center, is 
$(a\tau_0)^{1/3}$ \citep{Ada75,Neu90}; the physical distance corresponding to 
one optical depth at the line center is found by equating
$\Delta \tau_0 = 5.90\times10^{-12}n_{\rm H} \times \Delta l/\sqrt{\TD}  = 1$.
From this it follows that 
$\epsilon\tau_0 = (a\tau_0)^{1/3} N_{\rm H} {\sigma_{gr}}^{abs}$
where  ${\sigma_{gr}}^{abs} = 1.27\times10^{-21}~\pcc$ has been calculated
recently by \cite{WeiDra01} (their value for the albedo is 0.32). Several 
cases of lossy media are treated at length by \cite{Neu90} (analytically) and 
by \cite{Ada71a} (numerically).  The formalism discussed here reproduces 
the numerical results in the latter reference quite well but the grain 
model of \cite{WeiDra01} is about 30\% more absorptive 
than the worst case considered by Adams.  

In Figure 4 we show the scale factor f multiplying R/c as a function of
$\epsilon\tau_0$.  For a lossless medium, the exact analytic solution
yields f=0.909 at the line center and slab midplane(vs 0.993).  Losses are 
relatively unimportant 
in the intercloud gas over columns which are only a few pc long, $i.e.$
those which might be associated with individual cloud envelopes (the case 
where the size of the medium is comparable to the galactic scale 
height is considered in the next Section).  But the attenuation of \Lya\ 
radiation will be significant in any cool gas which absorbs strongly at 
$\lambda21$cm.

\subsection{Efficacy of the locally-generated \Lya\ photon field}

Figure 3 is a behind-the-scenes look at the excitation of the hyperfine
line as it occurs under the conditions used to produce Figures 1-2; 
it shows the y-values for collisions and \Lya\ excitation.   For the
purposes of the calculation, we took a characteristic size R = 1.5pc.
\Lya\ excitation would compete strongly with particle excitation in 
cooler gas were it not for the effects of dust absorption and the 
decrease in ionization fraction due to recombination on grains.  In warmer 
gas, the small physical size of the medium assumed for the calculation 
limits the effectiveness of \Lya\ excitation except where the gas ionizes 
fully.  

We conclude that there is no locally-generated source of excitation 
within individual clouds which will suffice to thermalize the spin 
temperature of warm H I, although the small attenuation in warm gas 
means that \Lya\ excitation can become important if galaxian scales are 
considered.  Because the Galaxy is a copious producer of \Lya\ photons,
the galactic \Lya\ radiation field will dominate excitation of the 
$\lambda$21cm line under some assumed conditions, as discussed below.

\section{\Lya\ excitation by Galactic radiation}

The space-averaged volume creation rate of \Lya\ photons near the galactic 
plane dwarfs that generated {\it in situ} in H I gas by large factors.  The 
galactic density of \Lya\ photons is not known with certainty but estimates 
are enough to force consideration of its importance in exciting the neutral
H I in warm media.

\cite{VacGar+96} have shown that the areal production rate of ionizing
photons from 429 O- and early-B stars within 2.5 kpc of the Sun is 
$\Psi_0 = 3.7\times10^7$ photons $ ~\pcc~\ps$.  Each ionizing photon should
produce of order one \Lya\ photon. Thus, over a cylinder of
height h = $\pm 50$ pc = 100 pc about the Galactic midplane, the 
mean volume production rate is 
$\dot n_{\alpha} \approx \Psi_0/h = 1.2\times10^{-13}~\pccc~\ps$.
This is far higher than that generated
solely by recombination in the warm neutral gas we modelled: the combination
of a long path and higher volume production rate would result in a
vastly increased line excitation by \Lya\ photons.
\footnote{This value of $\Psi_0$ is three times higher than that of 
\cite{DraSal78} who added the separate contributions of H II regions, 
recombination in diffuse hydrogen, and production in supernova 
remants (any such extra contributions should in fact be included).  
\cite{Ada71} cites an observational limit on the galactic \Lya\ flux due 
to \cite{ThoKra71} but this limit is consistent with large values of
$\Psi_0$ given recent estimates of the continuous absorption by dust.}

\begin{figure}
\psfig{figure=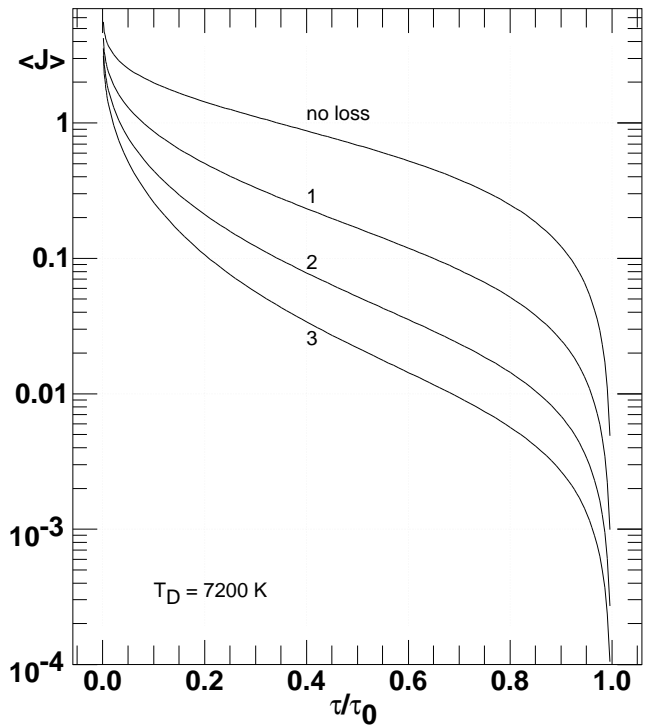,height=9.5cm}
\caption[]{Variation of the \Lya\ radiation field inside a slab of warm H I 
harboring a unit plane source $\delta(\tau)$ at the midplane, plotted against 
distance from the slab center $\tau/\tau_0$; see Section 6.1 of the text.  
Results for a lossless slab are shown, and for 
N$_{\rm H}$ = 1, 2, and $3 \times 10^{20}~\pcc$.}
\end{figure}

In Figure 5 we show the behaviour of \Tsp\ when a specified fraction
F$_\alpha$ of the galactic \Lya\ flux threads the gas assuming 
Nw$=10^{19}~\pcc$ and \TL = \TK.  Behaviour in the upper left panel of 
Figure 5 is identical to that in Figure 2.  Figure 5 
shows that even a very small fraction of the galactic \Lya\ radiation 
suffices to dominate the excitation.

Apparently, there is an abundant source of non-collisional excitation, 
if only it can be used.

\subsection{A two-phase ISM}

We consider first an idealized two-phase interstellar medium which fills
all space, and in which nearly all of the volume is occupied by warm intercloud
gas.  Following the discussion of this problem by \cite{Ada71a} we idealize 
the thin layer of young stars to be a plane source embedded in the much
thicker, warm H I layer, and apply an analytic radiative transfer solution
using the new value of the grain absorption cross-section from 
\cite{WeiDra01} as mentioned earlier.  At first, we neglect the fact 
that most photons which enter cold H I clouds will be absorbed by dust.

Figure 6 shows our evaluation of the analytic solution \citep{Har73,Neu90}
to this idealized problem, performed for a gas with \TK = \TD = 7200 K and 
various values of the column density in the H I layer (which determines
the loss due to continuous absorption by dust).  Although the total 
column density of H I looking out
from the midplane is about $3 \times 10^{20}~\pcc$ \citep{Lis83}, that
of the intercloud gas is somewhat less, of order half.  Figure 6 shows
that over most of the volume of this idealized model, the \Lya\
radiation field threading the intercloud gas is certainly not less
than 1\% of $\Psi_0$.  In this case, the spin temperature
in the intercloud gas at all but the highest galactic z-heights
would be close to equilibrium with the temperature of
the ambient galactic \Lya\ radiation -- \TK, \TD\ or what have you,
depending on how the photon field relaxes in its interactions with the gas.

Cool neutral gas packets dispersed throughout such a scattering medium
would clearly be threaded by a substantial fraction of the galactic
flux as well, raising the interesting possibility that the spin
temperatures measured in strongly absorbing gas are also influenced
by \Lya\ radiation.  A calculation analogous to that used to produce 
Figure 5 shows that \Lya\ excitation dominates if more than 0.1\% of the 
galactic radiation field threads cool gas.  In this case, \Tsp = \TD\ rather 
than \Tsp = \TK\ and typical turbulent velocity contributions to \TD\
will put a 30 - 120 K floor on measurements of \Tsp\ (see Section 6.3) 
{\it if} the turbulence is established on the scale of the \Lya\ scattering.

\begin{figure}
\psfig{figure=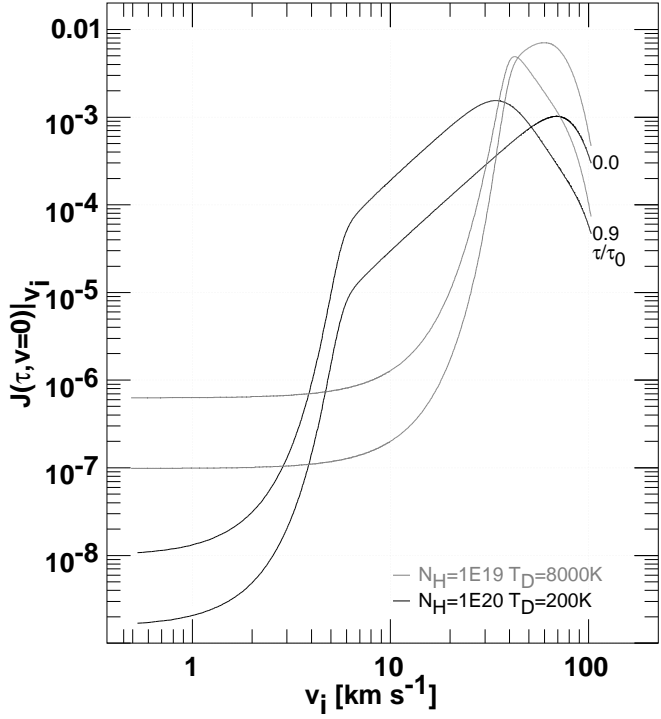,height=9.5cm}
\caption[]{Line-center \Lya\ mean intensity inside a lossless slab
in the presence of a half-unit monochromatic flux $\delta(v-v_i)/2$ 
incident on each face following the analytic solution of \cite{Neu90}.   
Two combinations of slab thickness and Doppler 
temperature are shown, representing cool and warm neutral gas.  For
each of these, the mean intensity is shown at the center and 90\% of the 
way from the mid-plane to the slab surface.  Over most of the interior,
the behaviour more closely resembles that at the midplane.}
\end{figure}

\subsection{A multi-phase ISM}

\cite{Neu91} pointed out that a multiphase ISM structure may foster escape 
of \Lya\ radiation and that individual neutral gas clouds -- warm or cool --
may not be threaded by much of 
the ambient \Lya\ flux.  If most of the volume of the ISM is in a 
contiguous hot phase having negligible neutral hydrogen or \Lya\ 
absorption, scattered radiation escapes relatively easily even when neutral
gas clouds have a very high surface covering factor.  Photons scatter off 
individual clouds as if they were particles of a gas having a Doppler 
temperature corresponding to the cloud-cloud velocity dispersion, 
6 \kms, appreciably narrowing the interstellar line profile if the 
photons are scattered sufficiently often before escape.  The crucial issue is 
the fraction and connectedness of the ISM occupied by very hot gas in 
which hydrogen is totally ionized, as opposed even to 6000-10000K H I 
intercloud gas which will still present a sizable opacity to \Lya\ radiation.

The fraction of incident photons which will be reflected back from the outer 
boundary of an isolated, discrete cloud is of order $1-1/(\tau_0\phi(x))$  
where x is the normalized frequency shift of the incident radiation from 
line center \citep{Neu90}: see Section 5.1 here.  An analytic solution 
to the problem of radiation incident
on a lossless slab is presented by \cite{Neu90} in his equations 2.19-2.22; 
calculation is considerably simplified when Neufeld's equation 2.22
\footnote{Note that the quantity plotted in Figure 2 of \cite{Neu90} is
$(a\tau_0)^{1/3}J(\tau_0,x)$, not $(a\tau_0)J(\tau_0,x)$.}
is
recast as 
$$ F(w,y) = 
 g \times \ln(1- 2 \exp(-\pi y) \cos(\pi w) +\exp(-2\pi y)),$$

$$  g\equiv \sqrt{6}/(16\pi^2). $$

In Figure 7 we show the mean intensity at the line center inside lossless 
slabs of warm and cool gas as the frequency shift of the attempted injection 
(expressed in \kms) varies; neglect of loss is appropriate for warm gas on 
the scale of individual clouds.  Any cool gas in the ISM is either embedded 
in a strongly-scattering intercloud gas or enrobed in warmer H I and its
internal radiation field could be much larger than that shown here;  the 
reflectivity calculations for bare cool gas are included only
to illustrate parameter sensitivities. 

If the ambient \Lya\ photon field is as narrow as the 6 \kms\ cloud-cloud 
velocity dispersion after multiple reflections, little will penetrate cool
gas and essentially none will enter the warm H I.
But if the \Lya\ photon field is appreciable at relative velocities of 
30-100 \kms\ with respect to the line center, the internal radiation field 
in warm neutral gas could be as high as 0.1-1\% of that which is incident.
The interstellar \Lya\ field would be quite broad in any model where the 
galactic layer harbors appreciable amounts of distributed warm H I 
\citep{Ada71} and \Lya\ photons will be injected into the ISM in the 
local line wings when they escape an H II region.  But the
ambient \Lya\ radiation field and its Doppler temperature 
are not easily specified in extreme three-phase models in which no
\Lya\ scattering occurs between individual clouds \citep{Neu91}.

\begin{figure}
\psfig{figure=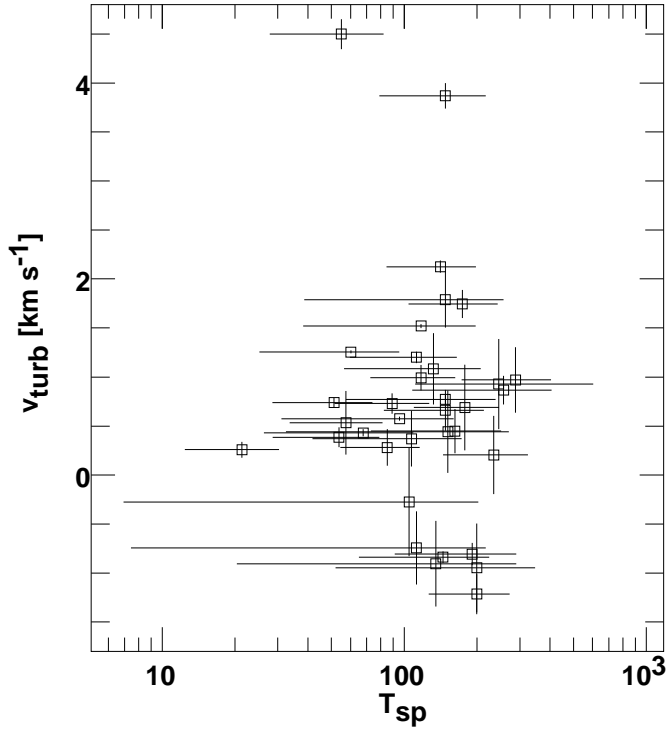,height=9.5cm}
\caption[]{Turbulent velocity contributions derived from
the spin and Doppler-temperature measurements of
\cite{PayTer+82,PaySal+83} using (\TD\ = \Tsub sp \ +121 \vturb$^2$)
for features having $\tau_0 \geq 0.04$.}
\end{figure}


\subsection{Non-thermal Doppler excitation}

For cool gas, where \Lya\ excitation would be important if
galactic radiation enters {\it via} the intercloud or warm neutral gas
which enrobes it, HI absorption features typically show a noticeable 
difference between \Tsp\ and \TD, implying that excitation of 
the hyperfine transition is {\it not} dominated by \Lya\ excitation
at a temperature other than \TK.

In Figure 8, we show the turbulent velocity contributions derived
from the measurements  of \cite{PayTer+82,PaySal+83}
where, for the stronger features ($\tau \geq\ 0.04$) and
excepting a few cases with very large quoted errors in \Tsp,
we have derived values of \vturb\ according to equation 3.
In those few cases where \TD\ $<$ \Tsp\ in the data, the value is shown 
negative.  The error estimates are ours, derived from analysis of the
errors quoted in the original data.

Clearly, most of the data are consistent with an added microturbulent 
contribution of order \vturb\ = 0.5-1.0 \kms, equivalent to adding
a 30 - 120 K turbulent contribution in quadrature with \TK\ to form
\TD.  Of course, if \Lya\ excitation is actually operating at a Doppler 
temperature above \TK, the measured spin temperatures are already somewhat 
inflated, and the turbulent velocities extracted from the comparison 
between \Tsp\ and \TD\ are only lower limits. Statistically, 
$\langle\Tsp/\TD\rangle = 0.56\pm0.26$ in this dataset.

Recent studies of stray-radiation-corrected Galactic emission profiles 
find evidence 
for three kinematic components having linewidths of 5, 13, and 31 \kms\
\citep{VerPer99}.  These correspond respectively to individual clouds 
(or to \TD = 550 K; see \cite{Tak67} and references therein), to the 
6 \kms\ cloud-cloud velocity dispersion 
(or  \TD = 3700 K), and to an anonymous component with \TD\ = 21000 K or a
combination of \TK\ = 8000K, v$_{\rm turb} = 10.4$ \kms.  Suggestions of
a pure intercloud component corresponding to 
\TK\ = 8000K, v$_{\rm turb} = 0$ \kms have a somewhat checkered history,
but seem now not to be clearly evident in the data.  This is perhaps not
too surprising given the panoply of phenomena which may influence an
observed linewidth on galactic scales. 

\section{Summary}

The major conclusions of this work may be summarized as follows:

Particle excitation does not thermalize warm (intercloud) H I in 
either the two- or three-phase models of the ISM.  In the absence 
of other sources of excitation, one would expect to see \Tsp\ 
$\approx 1000 - 5000$ K, from regions where \TK\ $\approx 5000-10000$ K 
(note that regions of higher \TK\ typically are at lower pressure in 
two-phase equilibrium and produce lower \Tsp), even when the contribution
of warm gas can be isolated.  The presence of a wide range of \Tsp\
in warm gas complicates the derivation of\Tsp\ in cool gas, which
is nearly always seen blended with warm gas along the line of sight.

In the two-phase model, at all but the highest galactic z-heights, hyperfine 
excitation in warm, neutral intercloud gas will generally be
dominated by the scattering of galactic \Lya\ photons produced by the 
ensemble of OB stars 
and supernova remnants, in which case the spin temperature will correspond 
to whatever Doppler temperature the \Lya\ photon field acquires while 
interacting with the interstellar gas.  Alternatively -- recalling comments 
in the Introduction -- excitation may be dominated locally by the 
\Lya\ radiation originating in or near particular objects.

In the three-phase model the ability of the galactic \Lya\ radiation field to 
penetrate individual H I clouds is somewhat problematic.  

In any multi-phase model where the warm neutral gas is threaded by a 
substantial galactic \Lya\ flux, excitation in the cool H I will be 
influenced by the \Lya\ photon field also.  

The excitation in cool H I clouds seems {\it not} to be totally dominated 
by  \Lya\ excitation at a temperature other than \TK, judging from the 
ubiquity of an apparent turbulent velocity contribution to the H I 
absorption linewidth, {\it i.e.} $\langle$\Tsp/\TD$\rangle \approx 0.56$.

The scattering calculations which will tell us the color temperature
of the interstellar \Lya\ radiation field remain to be performed and
may require more knowledge of the structure of the ISM than is currently
available to us.

\begin{acknowledgements}

The National Radio Astronomy Observatory is operated by AUI, Inc. under a
cooperative agreement with the US National Science Foundation.  I thank
David Neufeld and John Dickey for helpful comments on the manuscript
and Fred Schwab for providing a reference to a really good numerical 
representation of the Voigt function.

\end{acknowledgements}
 
\bibliographystyle{apj}
\bibliography{mnemonic,absorption}

\end{document}